\documentclass{emulateapj}
\usepackage{times,mathptm}
\usepackage{lscape}

\usepackage{epsf}
\usepackage{psfig}
\usepackage{lscape}

\def\etal{{et\,al.\,}}

\def\ltsima{$\; \buildrel < \over \sim \;$}
\def\simlt{\lower.5ex\hbox{\ltsima}}
\def\gtsima{$\; \buildrel > \over \sim \;$}
\def\simgt{\lower.5ex\hbox{\gtsima}}

\begin{document}

\title{X-ray Spectral Constraints for $z\approx$~2 Massive Galaxies:\\ The Identification of Reflection-Dominated Active Galactic Nuclei}


\author{D.~M. Alexander,\altaffilmark{1}
F.~E. Bauer,\altaffilmark{2,3}
W.~N. Brandt,\altaffilmark{4,5}
E. Daddi,\altaffilmark{6}
R.~C. Hickox,\altaffilmark{1}
B.~D. Lehmer,\altaffilmark{7,8}
B. Luo,\altaffilmark{4,5}
Y.~Q. Xue,\altaffilmark{4,5}
M.~Young,\altaffilmark{4,5}
A. Comastri,\altaffilmark{9}
A.~Del Moro,\altaffilmark{1}
A.~C. Fabian,\altaffilmark{10}
R. Gilli,\altaffilmark{9}
A.~D. Goulding,\altaffilmark{1,11}
V. Mainieri,\altaffilmark{12}
J.~R. Mullaney,\altaffilmark{1,6}
M.~Paolillo,\altaffilmark{13}
D.~A. Rafferty,\altaffilmark{4}
D.~P. Schneider,\altaffilmark{4}
O. Shemmer,\altaffilmark{14}
and C. Vignali\altaffilmark{15}
}

\affil{$^1$Department of Physics, Durham University, Durham DH1 3LE, UK}
\affil{$^2$Pontificia Universidad Cat\'{o}lica de Chile, Departamento de Astronom\'{\i}a y Astrof\'{\i}sica, Casilla 306, Santiago 22, Chile}
\affil{$^3$Space Science Institute, 4750 Walnut Street, Suite 205, Boulder, Colorado 80301}
\affil{$^4$Department of Astronomy and Astrophysics, 525 Davey Lab, Pennsylvania State University, University Park, PA 16802}
\affil{$^5$Institute for Gravitation and the Cosmos, The Pennsylvania State University, University Park, PA 16802}
\affil{$^6$Laboratoire AIM, CEA/DSM-CNRS-Universite Paris Diderot, Irfu/SAp, Orme des Merisiers, F-91191 Gif-sur-Yvette, France}
\affil{$^7$The Johns Hopkins University, Homewood Campus, Baltimore, MD 21218}
\affil{$^8$NASA Goddard Space Flight Centre, Code 662, Greenbelt, MD 20771}
\affil{$^9$INAF-Osservatorio Astronomico di Bologna, Via Ranzani 1, I-04127 Bologna, Italy}
\affil{$^{10}$Institute of Astronomy, Madingley Road, Cambridge CB3 0HA, UK}
\affil{$^{11}$Harvard-Smithsonian Center for Astrophysics, 60 Garden Street, Cambridge, MA 02138}
\affil{$^{12}$European Southern Observatory, Karl-Schwarzschild-Strasse 2, D-85748 Garching, Germany}
\affil{$^{13}$Dipartimento di Scienze Fisiche, Universita di Napoli ``Federico II,'' Complesso Universitario di Monte S. Angelo V. Cinthia, 9, I-80126, Napoli, Italy}
\affil{$^{14}$Department of Physics, University of North Texas, Denton, TX 76203}
\affil{$^{15}$Dipartimento di Astronomia, Universita degli Studi di Bologna, Via Ranzani 1, 40127 Bologna, Italy}

\shorttitle{REFLECTION-DOMINATED AGNS IN $z\approx$~2 MASSIVE GALAXIES}

\shortauthors{ALEXANDER ET AL.}

%
\begin{abstract} 
%

We use the 4~Ms {\it Chandra} Deep Field-South (CDF-S) survey to place
direct constraints on the ubiquity of $z\approx$~2 heavily obscured
AGNs in $K<22$ $BzK$-selected galaxies. Forty seven ($\approx$~21\%)
of the 222 $BzK$-selected galaxies in the central region of the CDF-S
are detected at X-ray energies, 11 ($\approx$~5\%) of which have hard
X-ray spectral slopes ($\Gamma\simlt1$) indicating the presence of
heavily obscured AGN activity ($N_{\rm
H}\simgt3\times10^{23}$~cm$^{-2}$). The other 36 X-ray detected $BzK$
galaxies appear to be relatively unobscured AGNs and starburst
galaxies; we use X-ray variability analyses over a rest-frame baseline
of $\approx$~3 years to further confirm the presence of AGN activity
in many of these systems. The majority (seven out of 11) of the
heavily obscured AGNs have excess infrared emission over that expected
from star formation (termed ``infrared-excess galaxies''). However, we
find that X-ray detected heavily obscured AGNs only comprise
$\approx$~25\% of the infrared-excess galaxy population, which is
otherwise composed of relatively unobscured AGNs and starburst
galaxies. We find that the typical \hbox{X-ray} spectrum of the
heavily obscured AGNs is better characterized by a pure reflection
model than an absorbed power-law model, suggesting extreme
Compton-thick absorption ($N_{\rm H}\simgt10^{24}$~cm$^{-2}$) in some
systems. We verify this result by producing a composite rest-frame
2--20~keV spectrum, which has a similar shape as a
reflection-dominated X-ray spectrum and reveals an emission feature at
rest-frame energy $\approx$~6.4~keV, likely to be due to Fe~K. These
heavily obscured AGNs are likely to be the distant analogs of the
reflection-dominated AGNs recently identified at $z\approx$~0 with
\hbox{$>10$~keV} observatories. On the basis of these analyses we
estimate the space density for typical (intrinsic X-ray luminosities
of $L_{\rm 2-10 keV}\simgt10^{43}$~erg~s$^{-1}$) heavily obscured and
Compton-thick AGNs at $z\approx$~2. Our space-density constraints are
conservative lower limits but they are already consistent with the
range of predictions from X-ray background models.

\end{abstract}

\keywords{galaxies: active --- galaxies: high-redshift --- infrared:
  galaxies --- X-rays: galaxies --- ultraviolet: galaxies}



%
\section{Introduction}
%

Deep X-ray surveys have provided a penetrating probe of Active
Galactic Nuclei (AGN) out to $z\approx$~5 (e.g.,\ Brandt \& Hasinger
2005; Silverman et~al. 2008; Brusa et~al. 2009; Brandt \& Alexander
2010), identifying obscured and unobscured AGN activity in a modest
fraction of the field-galaxy population ($\simgt$~5--10\%; e.g.,\
Lehmer \etal 2005, 2008; Xue \etal 2010). However, there is
overwhelming evidence that a large fraction of the heavily obscured
AGN population (\hbox{$N_{\rm H}\simgt3\times10^{23}$~cm$^{-2}$})
remains undetected in even the deepest X-ray surveys (e.g.,\ Worsley
\etal 2005; Hickox \& Markevitch 2006; Tozzi \etal 2006; Treister
\etal 2006; see \S1 of Alexander \etal 2008). Distant heavily obscured
AGNs are predicted by many theoretical models and simulations to
represent an important phase in the evolution of distant
dust-enshrouded galaxies, where the rapidly growing central
supermassive black hole (BH) is hidden from view (e.g.,\ Fabian 1999;
Granato \etal 2006; Hopkins \etal 2006). Therefore the identification
of the most heavily obscured AGNs could be more than just a
book-keeping exercise -- without having observations sensitive to
their identification we may miss a crucial BH growth phase.

Weak (faint or undetected) X-ray emission from luminous AGNs is likely
to be due to the presence of large amounts of dust/gas, sometimes
exceeding $N_{\rm H}\simgt10^{24}$~cm$^{-2}$ (i.e.,\ Compton thick;
e.g.,\ Matt \etal 2000; Comastri 2004; Della Ceca et~al. 2008; Murphy
\& Yaqoob 2009). Strong support for this statement comes from the
tight correlation between the optical and X-ray emission of unobscured
quasars (e.g.,\ Vignali, Brandt, \& Schneider 2003; Steffen
et~al. 2006; Gibson \etal 2008), which suggests that all luminous AGNs
are {\it intrinsically} bright at \hbox{X-ray} energies. Although weak
at X-ray energies, these heavily obscured AGNs should still be
detected in deep mid-to-far infrared (IR; rest-frame wavelength
$>2$~$\mu$m) observations due to the presence of dust heated by the
hidden AGN. Indeed, a number of studies have revealed large
populations of X-ray undetected IR-bright galaxies at $z\approx$~2,
which may host heavily obscured, potentially Compton thick, AGN
activity (e.g.,\ Daddi \etal 2007a; Fiore \etal 2008, 2009;
Georgantopoulos \etal 2008, 2011; Treister \etal 2009; Georgakakis
et~al. 2010). Many of these studies have employed X-ray stacking
analyses of X-ray undetected IR galaxies to identify hidden AGN
populations statistically, where the detection of a hard X-ray
spectral slope ($\Gamma\simlt1$) in the stacked data provides
compelling evidence for the presence of heavily obscured AGN activity
in at least a fraction of the stacked sources.\footnote{The spectral
slopes of X-ray emission from star-formation processes are typically
$\Gamma\simgt$~1 (e.g.,\ Kim \etal 1992; Ptak \etal 1999; Berghea
\etal 2008; Iwasawa \etal 2009).}  Under the assumption that all of
these X-ray undetected IR galaxies host Compton-thick AGN activity,
the implied space density of these hidden AGNs would exceed those of
the Compton-thin AGN population by a factor of $\simgt$~2, implying
that a much larger fraction of distant luminous AGNs are Compton thick
than found locally ($\simlt$~25--50\% of local AGNs appear to be
Compton thick; see Table~2 in Burlon et~al. 2010; Risaliti
et~al. 1999; Guainazzi et~al. 2005). These discoveries provide some
support for the hypothesis that the majority of distant BH growth was
more heavily obscured than that found locally (e.g.,\ La Franca \etal
2005; Treister \& Urry 2006; Hasinger 2008).

However, results from X-ray stacking analyses need to be treated with
caution since they only provide an average signal, leaving significant
uncertainties about the overall distribution of source
properties. Therefore, before strong conclusions can be derived from
these studies, at least two key questions need to be addressed: (1)
what fraction of the \hbox{X-ray} stacked signal is ``contaminated''
by star-forming galaxies, which are found to comprise at least a
fraction of the candidate heavily obscured AGN population (e.g.,\
Donley et~al. 2008; Murphy et~al. 2009; Fadda et~al. 2010)?  (2) what
fraction of the heavily obscured AGNs are absorbed by Compton thick
(as opposed to Compton thin) material? The most direct way to address
these questions is with deeper X-ray data, which will (1) reveal the
X-ray properties of individual IR-bright galaxies that were previously
contributing to the stacked \hbox{X-ray} signal, (2) allow for more
detailed X-ray spectral investigations of the X-ray detected IR-bright
galaxies to search for the signatures of heavily obscured and
Compton-thick absorption (e.g.,\ the identification of a strong
reflection component at rest-frame $>10$~keV; the detection of a high
equivalent width Fe~K emission line; e.g.,\ Matt \etal 1996, 2000;
Tozzi et~al. 2006; Georgantopoulos \etal 2009; Comastri \etal 2011;
Feruglio \etal 2011; see Murphy \& Yaqoob 2009),\footnote{We note that
high signal-to-noise ratio data is required to accurately distinguish
between the X-ray spectral properties of Compton-thin AGNs with
$N_{\rm H}\approx$~\hbox{(5--10)}~$\times10^{23}$~cm$^{-2}$ and
Compton-thick AGNs with $N_{\rm H}\simgt10^{24}$~cm$^{-2}$ (e.g.,\
Murphy \& Yaqoob 2009; Yaqoob \etal 2010). Therefore our adopted
definition of an Compton-thick AGN is not comprehensive but it is
conventional; see the MYTorus manual at www.mytorus.com for a detailed
review.} and (3) improve stacking constraints of X-ray undetected
populations. The presence of large amounts of absorption can also be
indirectly inferred from the identification of luminous AGN emission
lines and an IR-emitting hot-dust AGN continuum in X-ray weak systems
(e.g.,\ Bassani \etal 1999; Krabbe \etal 2001; Lutz \etal 2004;
Alexander \etal 2005b, 2008; Heckman \etal 2005; Gandhi \etal 2009;
Bauer \etal 2010; Donley \etal 2010; Gilli \etal 2010; Vignali \etal
2010; Goulding \etal 2011). Greater reliability in the identification
of heavily obscured and Compton-thick AGNs is made when considering
multiple diagnostics that cross-check each other, particularly those
that probe different AGN regions.

In this paper we use the deepest X-ray observations available (the
4~Ms {\it Chandra} exposure of the {\it Chandra} Deep Field-South;
CDF-S; Xue \etal 2011) to extend the analyses of Daddi \etal (2007a),
which employed X-ray stacking techniques to study X-ray undetected
IR-bright $z\approx$~2 galaxies in the shallower 1~Ms CDF-S
observations. Daddi \etal (2007b) utilized the $BzK$
photometric-selection technique (Daddi \etal 2004) to identify $K<22$
galaxies at $z\approx$~1.4--2.6 and classified objects based on the
ratio of mid-IR (24~$\mu$m) to extinction-corrected ultra-violet (UV;
rest-frame 1500~\AA) star-formation rates (SFRs). The $BzK$
photometric-selection technique provides an effective identification
of massive galaxies ($\approx10^{10}$--$10^{11}$~$M_{\odot}$; see
Daddi \etal 2007b; McCracken \etal 2010). $BzK$ galaxies with a
significant excess of IR emission over that predicted from the
extinction-corrected UV SFRs were classified as ``IR-excess' galaxies"
[log(SFR(mid-IR+UV)/SFR(UV,corr))$>0.5$] while $BzK$ galaxies with
comparable mid-IR and extinction-corrected UV SFRs were classified as
``IR-normal galaxies" [log(SFR(mid-IR+UV)/SFR(UV,corr))$\le0.5$]. From
stacking the X-ray data of the X-ray undetected galaxies, Daddi \etal
(2007a) obtained distinctly different X-ray spectral slopes for the
IR-excess galaxies ($\Gamma\approx$~0.9) and IR-normal galaxies
($\Gamma\approx$~1.8). The flat X-ray spectral slope found for the
IR-excess galaxies indicates that a fraction of the X-ray undetected
IR-excess galaxy population host heavily obscured AGN activity, some
of which may be Compton thick; by comparison, the stacked X-ray
emission from the X-ray undetected IR-normal galaxies is consistent
with that expected from star formation. Under the assumption that all
of the IR-excess galaxies are Compton-thick AGNs, Daddi \etal (2007a)
estimated a space density of $\Phi\approx2.6\times10^{-4}$~Mpc$^{-3}$
for $z\approx$~2 Compton-thick AGNs with $L_{\rm 2-10
keV}>10^{42}$--$10^{43}$~erg~s$^{-1}$. With the deeper {\it Chandra}
data from Xue \etal (2011) we can now better characterize the X-ray
properties of both the X-ray detected and X-ray undetected $BzK$
galaxies and improve constraints on the ubiquity of distant heavily
obscured and Compton-thick AGNs. We adopt
$H_{0}=71$~km~s$^{-1}$~Mpc$^{-1}$, $\Omega_{\rm M}=0.27$, and
$\Omega_{\Lambda}=0.73$ throughout. The Galactic absorption toward the
CDF-S region is $N_{\rm H}=$~8.8~$\times10^{19}$~cm$^{-2}$ (Stark
et~al. 1992). All given magnitudes are based on the Vega-magnitude
system.

%
\section{Data and Stacking Procedures}
%

\subsection{Galaxy sample}

We use an updated version of the $z\approx$~2 galaxy samples generated
by Daddi \etal (2007b) in the CDF-S field. Due to small refinements in
the optical--mid-IR photometry and revised redshift estimates, our
updated sample is slightly different to that used in Daddi \etal
(2007a,b). However, qualitatively, these samples are the same as those
used in Daddi \etal (2007a,b) and have the same global properties.

To provide a good compromise between excellent X-ray sensitivity and a
large number of galaxies, we have only explored the X-ray properties
of $BzK$ galaxies that lie within $5\farcm5$ of the average {\it
Chandra} aimpoint. Within this region there are 76 objects classified
as IR-excess galaxies and 146 objects classified as IR-normal
galaxies.

\subsection{X-ray matching}

We matched the parent $BzK$ galaxy sample of 222 objects to the 4~Ms
{\it Chandra} catalogs of Xue \etal (2011) using a $1\farcs5$ search
radius. In total, 47 $BzK$ galaxies have an X-ray counterpart, 28 of
which are classified as IR-excess galaxies and 19 of which are
classified as IR-normal galaxies. The median X-ray--$K$-band position
offset is $0\farcs4$, which agrees with the median uncertainty of the
X-ray source positions (which correspond to the 68\% confidence level;
see Fig.~6a and Eqn.~2 in Xue \etal 2011); on the basis of our
matching parameters we expect $\approx$~1.6 spurious matches.  Twenty
of these $BzK$ galaxies were detected in the 1~Ms CDF-S catalogs of
Alexander \etal (2003) and a further 27 are now detected in the 4~Ms
CDF-S catalogs of Xue \etal (2011). The properties of the \hbox{X-ray}
detected $BzK$ galaxies are presented in Table~1.

Thirty of the X-ray detected $BzK$ galaxies have spectroscopic
redshifts: 27 from optical spectroscopy (predominantly from VLT
observations with the FORS1, FORS2, and VIMOS instruments; Appenzeller
et~al. 1998; Le F{\`e}vre et~al. 2003) and 3 from {\it Spitzer}-IRS
mid-IR spectroscopy; see Table~1 for the spectroscopic-redshift
references. The median spectroscopic redshift is $z_{\rm
spec}=1.78\pm0.35$.\footnote{The error on the median is the median
absolute deviation (MAD), which is a robust estimator of the spread of
the sample: ${\rm MAD}=1.48\times{\rm median}(|x-{\rm median}(x)|)$;
see \S1.2 of Maronna \etal (2006).} The other 17 X-ray detected $BzK$
galaxies have photometric redshifts and a median redshift of $z_{\rm
photo}=2.18\pm0.50$; see Table~1 for the photometric-redshift
references. The median absolute uncertainty between the photometric
and spectroscopic redshifts for the X-ray sources with spectroscopic
redshifts is $|\Delta{z}|/(1+{z_{\rm spec}})$~=~0.02, where
$\Delta{z}$~=~$z_{\rm photo}$~--~$z_{\rm spec}$; we get comparable
results if we use the Cardamone \etal (2010) photometric-redshift
catalog ($|\Delta{z}|/(1+{z_{\rm spec}})$~=~0.02).

%
%
\begin{figure}
\centerline{\includegraphics[angle=0,width=9.0cm]{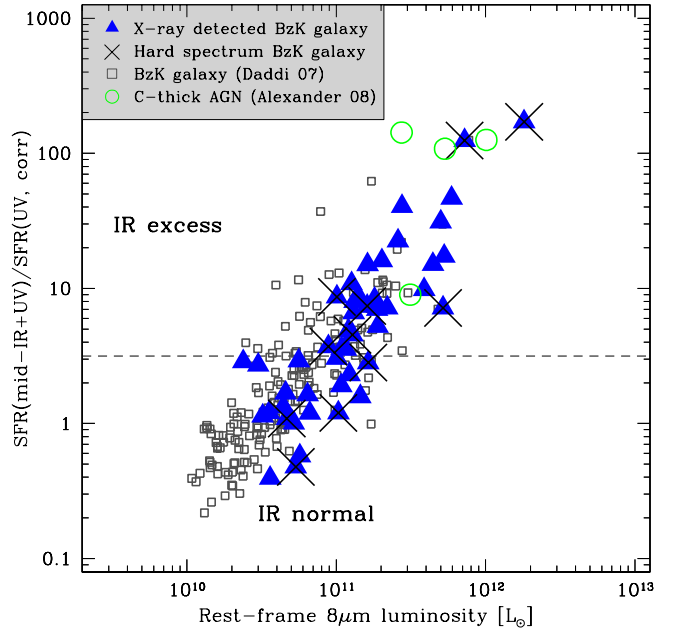}}
\caption{Ratio of star-formation rates (SFRs; mid-IR and
  extinction-corrected UV) for the $z\approx$~2 galaxies studied by
  Daddi \etal (2007a,b) versus rest-frame 8~$\mu$m
  luminosity. Triangles correspond to X-ray detected $BzK$ galaxies,
  crosses indicate the $BzK$ galaxies with hard X-ray spectral slopes
  (heavily obscured AGNs; see Fig.~2), open squares correspond to the
  X-ray undetected $BzK$ galaxies, and the open circles correspond to
  the spectroscopically identified (optical and mid-IR wavelengths)
  Compton-thick AGNs at $z\approx$~2 from Alexander \etal (2008) in
  the 2~Ms {\it Chandra} Deep Field-North survey (Alexander \etal
  2003). The threshold between IR excess and IR normal galaxies
  defined by Daddi \etal (2007a,b) is indicated by the dashed line.}
\end{figure}

\subsection{X-ray spectroscopy}

We extracted and analysed the X-ray spectra of the X-ray detected
$BzK$ galaxies to provide greater insight into their intrinsic
properties. The X-ray spectra were extracted using {\sc acis\_extract}
(Broos et~al. 2010) as part of the \hbox{X-ray} catalog construction
in Xue \etal (2011); see \S3.2 of Xue \etal (2011) for more details.

Due to the limited counting statistics for the majority of the heavily
obscured AGNs ($\simlt$~80 counts in the \hbox{0.5--8~keV} band; see
Table~2), which are the primary focus of this paper, the X-ray
spectral analyses were predominantly performed using the $C$-statistic
(Cash 1979). The $C$-statistic is calculated on the unbinned data and
is therefore ideally suited to low-count sources (e.g.,\ Nousek \&
Shue 1989). However, to provide consistency checks on these results,
we also performed \hbox{X-ray} spectral analyses of the brightest
X-ray sources ($>$~200 counts in the 0.5--8~keV band) using $\chi^2$
statistics; in these analyses we grouped the X-ray data into 20 counts
per bin. All fit parameter uncertainties are quoted at the 90\%
confidence level (Avni 1976).

\subsection{X-ray variability}

We analysed the X-ray variability of the X-ray detected \emph{BzK}
galaxies to look for nuclear activity. The CDF-S observations were
split into four epochs, each approximately 1~Ms long: 2000, 2007,
2010a (March - May), and 2010b (May - July). Within each epoch, the
observations were merged and photometry was measured using
\textsc{acis\_extract} (Broos et al. 2010), as described in detail in
Xue et al. (2011).

We apply a $\chi^2$ test to determine if a source is variable by
comparing the variability observed between observations to that
expected from Poisson statistics.  The test statistic will follow a
$\chi^2$ distribution except at low count rates, where the errors are
larger than expected from a Gaussian distribution. In the low-count
regime, the test statistic is smaller than expected and does not
follow the $\chi^2$ distribution. We construct a Monte Carlo
simulation to determine what distribution the test statistic should
follow for each source, following the procedure of Paolillo et
al. (2004). The observed test statistic is compared to the simulated
distribution to determine the probability ($P_{\chi^2}$) that the
observed variability is due to Poisson noise. The $\chi^2$ and
$P_{\chi^2}$ values are listed in Table~2. A source is considered
variable if it has $>20$ counts in the 0.5--8~keV band and
$P_{\chi^2}$ $<$ 5\%.

The normalized excess variance ($\sigma^2_{rms}$; Nandra et al. 1997)
measures how strongly each source varies in excess of measurement
error.  Since it is more likely that a variable galaxy harbors an AGN
if the observed variability exceeds that expected from a population of
X-ray binaries, we estimate the amount of variability expected from
high-mass X-ray binaries (HMXBs). HMXB variability can be estimated
from a galaxy's star formation rate (see Fig.~8 of Gilfanov et
al. 2004). To use this relation we adopt $\sigma^2_{rms}$~=~0.09 as
the typical variability for an individual X-ray binary (equivalent to
30\% fractional rms; see \S6 of Gilfanov 2010). This value is the
maximum variability expected for a HMXB population. Variability
strength will increase with the length of the timescale over which it
is measured (e.g.,\ Nandra \etal 1997), so adopting the maximum HMXB
variability is appropriate for CDF-S sources at $z\approx$~2, where
rest-frame timescales are $\approx$~3~years. The SFR of the \emph{BzK}
galaxies was estimated from their UV and IR luminosities (Xue et
al. 2010). Applying the $\sigma_{rms}^2$-SFR relations (Gilfanov et
al. 2004), we calculate an upper limit to the HMXB contribution to the
variability, reported as $\sigma_{\text{HMXB}}^2$ in Table~2. A source
is considered to be an X-ray variable AGN if it is found to be
variable, as prescribed above, and also has excess variability over
that expected from HMXBs. On the basis of these criteria, 13 of the 27
sources with $>20$ counts in the 0.5--8~keV band are found to be
variable AGN.

\subsection{X-ray stacking analyses}

We used X-ray stacking analyses to constrain the average X-ray
properties of the X-ray undetected $BzK$ galaxy populations. In our
X-ray stacking analyses we adopted the procedure of Lehmer \etal
(2008), which takes a different approach from the Worsley \etal (2005)
method used by Daddi \etal (2007a). Both procedures stack the X-ray
data of the selected sources but Lehmer \etal (2008) determine the
background counts using large source-free apertures local to each
source while Worsley \etal (2005) determine the background counts from
a large number of randomly placed apertures around the source (i.e.,\
a Monte-Carlo approach). We tested both procedures on our datasets and
achieved statistically consistent results. The major advantage of the
Lehmer \etal (2008) approach over that of Worsley \etal (2005) is
computational speed. In the stacking analyses, we used a fixed
aperture of $1\farcs5$ radius and determined background counts in
$25^{\prime\prime}\times25^{\prime\prime}$ source-free regions local
to each source. We applied aperture corrections to the net stacked
count rates following \S4.2 in Lehmer \etal (2008).

A general concern in stacking analyses is that a few sources can
dominate the stacked signal. To guard against this, we randomly
selected 80\% of the objects in each stacking analysis sample and
stacked their properties. For each sample we performed this procedure
10,000 times, randomly selecting 80\% of the objects for each
iteration, to generate a distribution of the stacked properties. We
found that the overall properties obtained from the stacking analyses
were in good agreement with the overall properties found from the
stacking analysis trials, indicating that bright sources do not
dominate the stacked signal.

%
\section{Analyses and Results}
%

%
%
\begin{figure}
\centerline{\includegraphics[angle=0,width=9.0cm]{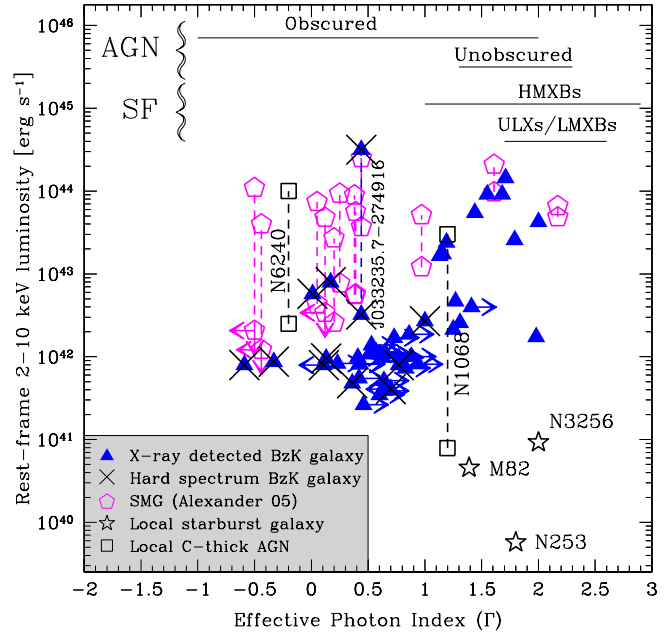}}
\caption{Rest-frame 2-10 keV luminosity versus X-ray spectral slope
  ($\Gamma$ is in the observed-frame 0.5--8~keV, typically rest-frame
  1.5--24~keV). The properties of $z\approx$~2 SMGs hosting AGN
  activity (open pentagons; from Alexander \etal 2005a) and X-ray
  detected $BzK$ galaxies (filled triangles; identified here) are
  compared to well-studied local starbursts (open stars) and local
  Compton-thick AGNs (open squares). The properties of the local
  starburst galaxies and Compton-thick AGNs are taken from Ranalli
  \etal (2003), Matt \etal (1997), and Vignati \etal (1999). The
  vertical dashed lines indicate the difference between the observed
  X-ray luminosity and the absorption-corrected luminosity for the
  SMGs and Compton-thick AGNs (from Matt \etal 1997; Vignati \etal
  1999; Alexander \etal 2005a; Feruglio et~al. 2011); the intrinsic
  X-ray luminosity constraint for J033235.7--274916 is from Feruglio
  et~al. (2011), which is one of the heavily obscured AGNs in our
  sample. The range of X-ray spectral slopes found for typical AGNs
  and star-forming galaxies are illustrated (horizontal solid lines;
  constraints derived from Kim \etal 1992; Nandra \& Pounds 1994;
  Maiolino \etal 1998; Ptak \etal 1999; Berghea \etal 2008). Eleven of
  the X-ray detected $BzK$ galaxies have flat X-ray spectral slopes
  ($\Gamma\simlt1$) and are classifed as heavily obscured AGNs
  (crosses).}
\end{figure}

%
%
\begin{figure}
\centerline{\includegraphics[angle=0,width=9.0cm]{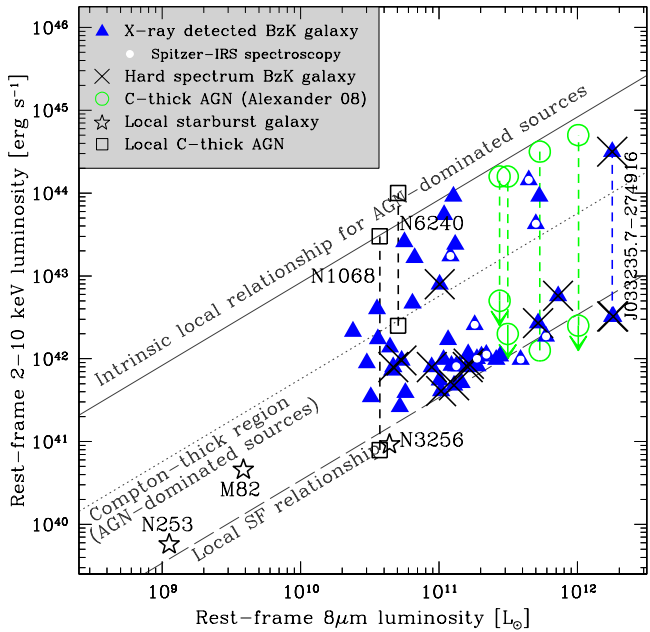}}
\caption{Rest-frame 2-10 keV luminosity versus rest-frame 8~$\mu$m
  luminosity. The symbols have the same meaning as in Figs.~1 \& 2; in
  addition, the small filled circles show the X-ray detected $BzK$
  galaxies with {\it Spitzer}-IRS spectroscopy. The X-ray--8~$\mu$m
  luminosity ratio for local starburst galaxies (long-dashed line) is
  taken from the X-ray--12~$\mu$m luminosity ratio of Krabbe \etal
  (2001) and converted to 8~$\mu$m assuming the M~82 spectral energy
  distribution. The intrinsic X-ray--8~$\mu$m luminosity ratio for
  local AGNs (solid line) is taken from Lutz \etal (2004) and
  converted to 8~$\mu$m, assuming the AGN-dominated galaxy NGC~1068;
  the dotted line indicates the observed X-ray--8~$\mu$m luminosity
  ratio predicted for Compton-thick AGNs. The rest-frame 8~$\mu$m
  luminosity for the X-ray detected $BzK$ galaxies is calculated from
  the 24~$\mu$m flux density, with small $K$-corrections applied (see
  Daddi \etal 2007a), while the rest-frame 8~$\mu$m luminosity for the
  local starburst galaxies is calculated using the mid-IR spectroscopy
  of Rigopoulou \etal (1999) and Lutz \etal (2003). The vertical
  dashed lines indicate the difference between the observed and
  intrinsic X-ray luminosity for the $z\approx$~2 Compton-thick AGNs
  from Alexander et~al. (2008) and Feruglio et~al. (2011); the
  intrinsic X-ray luminosity constraint for J033235.7--274916 is from
  Feruglio et~al. (2011), which is one of the heavily obscured AGNs in
  our sample.}
\end{figure}

We focus our analyses toward (1) characterizing the \hbox{X-ray}
spectral and variability properties of X-ray detected $BzK$ galaxies
to identify the presence of AGN activity (see \S3.1), and (2)
performing X-ray stacking analyses of the X-ray undetected $BzK$
galaxies in the 4~Ms CDF-S observations (see \S3.2). With the results
of these investigations we re-evaluate estimates of the space density
of distant heavily obscured and Compton-thick AGNs (see \S3.3).

\subsection{X-ray detected BzK galaxies}

In Fig.~1 we compare the properties of the X-ray detected $BzK$
galaxies to the overall $BzK$ galaxy population. The \hbox{X-ray}
detected $BzK$ galaxies cover a wide range in SFR ratio
(log(SFR(mid-IR+UV)/SFR(UV,corr)) and rest-frame 8~$\mu$m
luminosity. However, when compared to the X-ray undetected $BzK$
galaxies, the X-ray detected systems have characteristically higher
median SFR ratios and rest-frame 8~$\mu$m luminosities (X-ray
detected: SFR ratio of $\approx4.5\pm4.9$ and log($L_{\rm 8~\mu
m}/L_{\odot})\approx11.1\pm0.4$; \hbox{X-ray} undetected: SFR ratio of
$\approx1.4\pm1.4$ and log($L_{\rm 8~\mu
m}/L_{\odot})\approx10.6\pm0.4$).\footnote{$L_{\rm 8\mu
m}\simgt10^{11}$~$L_{\odot}$ is comparable to $L_{\rm
IR}\simgt10^{12}$~$L_{\odot}$ for the spectral energy distribution
corrections adopted in Daddi \etal (2007b).} Overall, $\approx$~37\%
of the IR-excess galaxy population and $\approx$~13\% of the IR-normal
galaxy population are now detected in the 4~Ms {\it Chandra} exposure,
indicating a close connection between the production of X-ray emission
and the presence of excess (or luminous) IR emission. Indeed, the
X-ray detected fraction rises as a function of rest-frame 8~$\mu$m
luminosity: $\approx$~3\%, $\approx$~14\%, and $\approx$~51\% of the
log($L_{\rm 8~\mu m}/L_{\odot})=$~10.0--10.5, log($L_{\rm 8~\mu
m}/L_{\odot})=$~10.5--11.0, and log($L_{\rm 8~\mu m}/L_{\odot})>$~11.0
systems are detected at X-ray energies, respectively. Given these
results, X-ray observations an order of magnitude deeper than those
obtained here (only likely to be attainable with the next generation
of X-ray observatories; e.g.,\ {\it Generation-X}; Wolk et~al. 2008)
are required to individually detect X-ray emission from the majority
of the lowest-luminosity systems.

\subsubsection{Classification of the X-ray emission}

In Fig.~2 we plot the rest-frame 2--10~keV luminosity versus X-ray
spectral slope of the X-ray detected $BzK$ galaxies and compare them
to well-studied local starburst galaxies, Compton-thick AGNs, and
$z\approx$~2 submillimeter-emitting galaxies (SMGs) hosting AGN
activity; the rest-frame \hbox{2--10~keV} luminosities are calculated
from the observed-frame 0.5--2~keV fluxes assuming $\Gamma=1.8$ for
the small K corrections. Eleven ($\approx$~23\%) of the 47 X-ray
detected $BzK$ galaxies have flat X-ray spectral slopes with
$\Gamma\simlt1$ (eight have $\Gamma\simlt0.5$) and are classified here
as ``heavily obscured AGNs''. On the basis of the X-ray properties of
the $z\approx$~2 SMGs hosting AGN activity (Alexander \etal 2005a),
the flat X-ray spectral slopes for these heavily obscured AGNs suggest
absorbing column densities of $N_{\rm
H}\simgt3\times10^{23}$~cm$^{-2}$ and some may be Compton thick; see
Fig.~2. The absorption corrections for such heavily obscured AGNs in
the rest-frame \hbox{2--10~keV} band are large and would imply
intrinsic luminosities of $L_{\rm 2-10
keV}\simgt10^{43}$~erg~s$^{-1}$. Six of these heavily obscured AGNs
are not detected in the 1~Ms CDF-S data used in Daddi \etal (2007a).

On the basis of the X-ray luminosity and X-ray spectral slope, the
X-ray emission from the majority of the other 36 X-ray detected $BzK$
galaxies is likely to be due to either relatively unobscured AGN
activity ($N_{\rm H}\simlt10^{22}$--$10^{23}$~cm$^{-2}$) or star
formation. Nine of these objects have $L_{\rm
X}>10^{43}$~erg~s$^{-1}$, five of which are identified as AGNs based
on optical spectroscopy (see Table~1), and are classified here as
``luminous AGNs''. The other 27 X-ray detected $BzK$ galaxies are
likely to be lower-luminosity AGNs or X-ray luminous starbursts and
are classified here as ``low-luminosity X-ray systems''; as shown in
Fig.~3 and \S3.1.2, the X-ray--IR luminosity ratios of these systems
are also similar to those expected for starburst galaxies or
$z\approx$~2 AGNs with $L_{\rm
X}\approx10^{42}$--$10^{43}$~erg~s$^{-1}$ (e.g.,\ Krabbe \etal 2001;
Mullaney \etal 2010; Shao \etal 2010). Twenty one of these
low-luminosity \hbox{X-ray} systems are not detected in the 1~Ms CDF-S
data used in Daddi \etal (2007a).

Four of the 27 low-luminosity X-ray systems are detected in both the
2--8~keV and 0.5--2~keV bands and their X-ray spectral slopes are
$\Gamma\approx$~1.2--2.0. The other 23 low-luminosity X-ray systems
are undetected at 2--8~keV and we can only provide accurate
constraints on their X-ray spectral slopes using X-ray stacking
analyses. Stacking the X-ray data for these 23 systems, following the
procedure outlined in \S2.5, we obtain significant detections in both
the 2--8~keV (S/N~=~8.3) and 0.5--2~keV bands (S/N~=~29.7), which
corresponds to an average X-ray spectral slope of
$\Gamma=1.5\pm0.1$. The comparatively steep X-ray spectral slope from
this population could be due to either star formation or AGN activity
and does not provide significant new insight into the composition of
the low-luminosity X-ray systems.

We can further characterize the X-ray detected $BzK$ galaxies using
X-ray variability analyses. The identification of significant X-ray
variability over that expected from star-formation processes will
indicate the presence of an AGN; see \S2.4. Overall, we find that 13
($\approx$~48\%) of the 27 X-ray detected $BzK$ galaxies with
reasonable-quality X-ray data ($>20$ X-ray counts) show excess
variability over that expected from star-formation processes; the
sources show variability by factors of $\approx$~1.4--4.3 (see
Table~2). Nine of these 13 systems had already been classified as
AGNs: six are luminous AGNs and four are heavily obscured
AGNs. However, three of the variable sources are classified as
low-luminosity X-ray systems ($\approx$~40\% of those with
reasonable-quality X-ray data), unambiguously identifying the presence
of AGNs in at least a fraction of the low-luminosity X-ray systems.

\subsubsection{X-ray--infrared properties}

In Fig.~3 we plot the rest-frame 2--10~keV luminosity versus
rest-frame 8~$\mu$m luminosity of the X-ray detected $BzK$ galaxies
and compare them to well-studied local starburst galaxies and
Compton-thick AGNs. This figure can help characterize the
X-ray--8~$\mu$m luminosity ratio and provide constraints on the
intrinsic luminosity of the heavily obscured AGNs. For example, under
the assumption that the 8~$\mu$m emission is dominated by AGN
activity, the X-ray--8~$\mu$m luminosity ratios of the heavily
obscured AGNs suggest that they have intrinsic X-ray luminosities of
$L_{\rm 2-10 keV}\approx3\times10^{43}$--$10^{45}$~erg~s$^{-1}$ (see
Fig.~3).  The presence of absorption at 8~$\mu$m, as predicted by
radiative-transfer modeling of clumpy AGN obscuration (e.g.,\ Nenkova
\etal 2008), will increase these instrinsic X-ray luminosity
estimates; however, see Lutz \etal (2004) and Gandhi \etal (2009) for
observational constraints suggesting that obscured AGNs do not
typically suffer significant nuclear absorption at infared
wavelengths. Under the assumption that the 8~$\mu$m emission is AGN
dominated, the X-ray--8~$\mu$m luminosity ratios for all of the
heavily obscured AGNs, except for J033222.5--274603, are also
consistent with those expected for Compton-thick AGNs.

However, the interpretation of the X-ray--8~$\mu$m luminosity ratio is
complicated by the absence of mid-IR spectroscopy for the majority of
the X-ray detected $BzK$ galaxies, which would directly measure the
contributions from star formation and AGN activity at rest-frame
8~$\mu$m. Nine of the X-ray detected $BzK$ galaxies have {\it
Spitzer}-IRS spectroscopy, three of which are classified as X-ray
luminous AGNs and six of which are classified as low-luminosity X-ray
sources; see Table~1 and Fig.~3. With the exception of the luminous
AGN J033237.7--275212 (which is AGN-dominated at 8~$\mu$m; Donley
et~al. 2010), all of the other 8 systems are star-formation dominated
at 8~$\mu$m, including two of the luminous AGNs (Teplitz et~al. 2007;
Fadda et~al. 2010). If the heavily obscured AGNs are the absorbed
counterparts of the luminous AGNs, as suggested by the X-ray spectral
analyses (see \S3.1.3), then we would expect many of them to also be
star-formation dominated at 8~$\mu$m and, therefore, the range of
intrinsic X-ray luminosities estimated above ($L_{\rm 2-10
keV}\approx3\times10^{43}$--$10^{45}$~erg~s$^{-1}$) are upper limits;
see \S3.1.4 for further estimates of the intrinsic X-ray luminosities.

The majority of the heavily obscured AGNs are identified as IR-excess
galaxies (seven of the 11 systems); see Fig.~1. However, the majority
of the X-ray detected IR-excess galaxies are not heavily obscured
AGNs: only 7 ($\approx$~25\%) of the 28 X-ray detected systems are
heavily obscured AGNs, 7 ($\approx$~25\%) are luminous AGNs, and 14
($\approx$~50\%) are low-luminosity X-ray systems (one of which has
been found to be an X-ray variable AGN: J033246.8--275120; see
Tables~1--2). This shows that the IR-excess galaxy population is
heterogenous, in qualitative agreement with several studies of
IR-excess galaxies (e.g.,\ Teplitz \etal 2007; Alexander \etal 2008;
Murphy \etal 2009; Fadda \etal 2010; Georgakakis \etal 2010;
Georgantopoulos \etal 2011).

%
%
\begin{figure}
\centerline{\includegraphics[angle=-90,width=9.0cm]{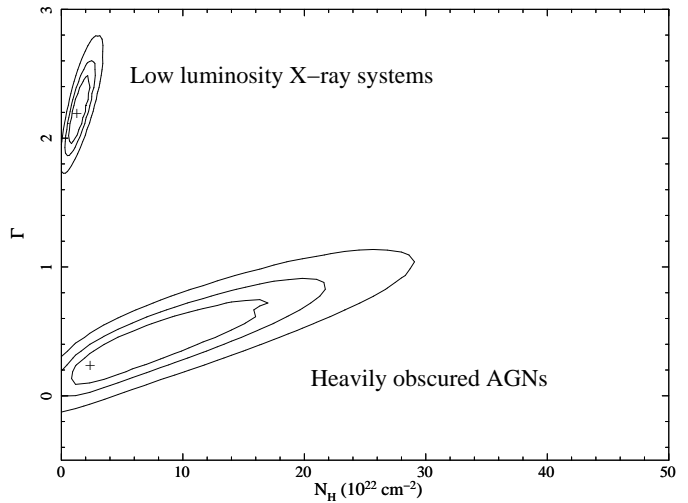}}
\caption{Best-fit parameters ($\Gamma$ versus $N_{\rm H}$) from
  jointly fitting the \hbox{X-ray} spectra of the heavily obscured
  AGNs and low-luminosity X-ray systems with an absorbed power-law
  model. The contours refer to the 68\%, 95\% and 99\% confidence
  limits and the crosses indicate the best-fitting parameters. The
  best-fitting $\Gamma$ for the heavily obscured AGNs is inconsistent
  with the intrinsic X-ray spectral slope found for AGN activity
  ($\Gamma\approx$~1.3--2.5; see Fig.~2), suggesting that the observed
  X-ray emission for the overall sample is not well characterized by
  absorbed power-law emission.}
\end{figure}

%
%
\begin{figure*}
\centerline{\includegraphics[angle=0,width=12.0cm]{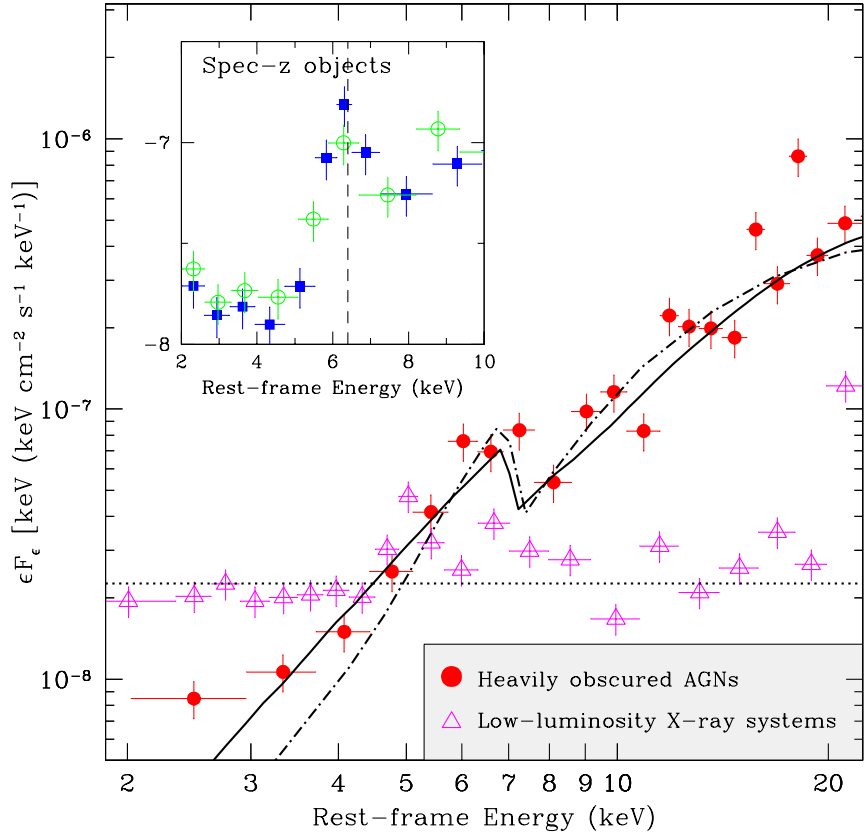}}
\caption{Flux density versus rest-frame energy showing the composite
  rest-frame 2--20~keV spectra for the heavily obscured AGNs (filled
  circles; all objects except the X-ray bright AGN J033222.5--274603)
  and low-luminosity X-ray systems (open triangles) as compared to an
  unabsorbed power-law model (dotted line; $\Gamma=2.0$), a pure
  reflection model (solid curve; $\Gamma=1.7$), and the best-fitting
  model to the reflection-dominated $z\approx$~0 AGN {\it
  Swift}~J0601.9--8636 (dot-dashed curve; Ueda \etal 2007); see
  \S3.1.4. The inset panel shows the stacked X-ray spectra at
  rest-frame 2--10~keV of the heavily obscured AGNs with spectroscopic
  redshifts (filled squares: all objects; open circles: all objects
  except J033235.7--274916, which has been individually identified
  with Fe~K$\alpha$ emission; Feruglio et~al. 2011); the dashed line
  shows the expected rest-frame energy of Fe~K$\alpha$. The properties
  of the heavily obscured AGNs are consistent with those expected for
  reflection-dominated systems and $\approx$~10--50\% are likely to be
  Compton-thick AGNs (see \S3.1.4).}
\end{figure*}

\subsubsection{X-ray spectral analyses}

To gain more insight into the intrinsic AGN properties (e.g.,\ $N_{\rm
 H}$, $\Gamma$, reflection components, Fe~K emission) of the X-ray
 detected $BzK$ galaxies we fitted the X-ray data using physically
 motivated AGN models. Our main focus here is to constrain the X-ray
 spectral properties of the 11 heavily obscured AGNs to identify any
 potential Compton-thick AGN signatures (dominant reflection
 component; strong Fe~K emission).  However, we also explored the
 X-ray spectral properties of all of the X-ray detected $BzK$ galaxies
 to provide constraints on key spectral parameters (e.g.,\ the
 intrinsic \hbox{X-ray} spectral slope; $\Gamma$) and search for
 further heavily obscured AGNs not identified using the simple X-ray
 spectral slope criteria. We extracted the X-ray spectra of each
 source following \S2.3 and initially fitted each X-ray spectrum over
 the observed-frame 0.5--8~keV energy band with an absorbed power-law
 model (the model components are {\sc wabs*zwabs*pow} in {\sc xspec})
 using the $C$ statistic (Cash 1979).

We first focus on the results obtained for the nine luminous AGNs
($L_{\rm X}>10^{43}$~erg~s$^{-1}$) since the good photon statistics
($\approx$~220--4600 net counts; mean of $\approx$~1600 net counts)
provide accurate constraints on the intrinsic X-ray spectral slope and
the presence of any absorption. The individual best-fitting parameters
for the luminous AGNs are consistent with those expected for
relatively unobscured AGNs ($\Gamma\approx$~1.8 and $N_{\rm
H}<10^{23}$~cm$^{-2}$). To provide tighter overall constraints we also
performed joint spectral fitting for all of the sources, which
determines the best-fitting $\Gamma$ and $N_{\rm H}$ for the whole
sample. Jointly fitting $\Gamma$ and $N_{\rm H}$ but leaving the
normalisation of each source to vary, we obtained
$\Gamma=1.71^{+0.03}_{-0.04}$ with low intrinsic absorption ($N_{\rm
H}=(0.65^{+0.06}_{-0.12})\times10^{22}$~cm$^{-2}$); the best-fitting
parameters obtained from $\chi^2$ fitting
($\Gamma=1.78^{+0.04}_{-0.04}$ and $N_{\rm
H}=(0.79^{+0.15}_{-0.10})\times10^{22}$~cm$^{-2}$) are statistically
consistent with those obtained using the $C$ statistic. These
constraints can help interpret the X-ray spectral properties of the
heavily obscured AGNs. The results obtained for the 27 low-luminosity
X-ray systems are similar to those obtained for the luminous AGNs but
with considerably larger uncertainties ($\approx$~3--180 net counts;
mean of $\approx$~30 net counts); the best-fitting parameters from
jointly fitting the X-ray spectra are $\Gamma=2.22^{+0.23}_{-0.29}$
with $N_{\rm H}=(1.46^{+1.01}_{-0.47})\times10^{22}$~cm$^{-2}$ (see
Fig.~4).

We now focus on the results obtained for the 11 heavily obscured AGNs.
The only heavily obscured AGN that is bright enough for
reasonable-quality individual X-ray spectral constraints is
J033222.5--274603 ($\approx$~560 net counts), which has
$\Gamma=0.93^{+0.28}_{-0.26}$ and $N_{\rm
H}=(0.82^{+0.41}_{-0.34})\times10^{23}$~cm$^{-2}$; the best-fitting
parameters obtained from $\chi^2$ fitting are
$\Gamma=1.03^{+0.26}_{-0.20}$ and $N_{\rm
H}=(1.80^{+0.58}_{-0.42})\times10^{23}$~cm$^{-2}$, with a reduced
$\chi^2$ of 1.13 for 24 degrees of freedom. The absorbing column
density of this object suggests that it is Compton thin; however, the
best-fitting X-ray spectral slope is flat. Examination of the
residuals shows that the model significantly deviates from the data at
observed-frame $<1.5$~keV. Fitting the X-ray data between
observed-frame 1.5--8~keV gives $\Gamma=1.40^{+0.30}_{-0.64}$ and
$N_{\rm H}=(2.05^{+0.97}_{-1.81})\times10^{23}$~cm$^{-2}$; the
best-fitting parameters obtained from $\chi^2$ fitting are
$\Gamma=1.59^{+0.52}_{-0.36}$ and $N_{\rm
H}=(2.68^{+1.66}_{-1.55})\times10^{23}$~cm$^{-2}$, with a reduced
$\chi^2$ of 1.03 for 20 degrees of freedom. These best-fitting
parameters now provide a better characterisation of the data and are
consistent with that expected for an obscured Compton-thin AGN. The
X-ray--8~$\mu$m luminosity ratio of $\approx$~0.02 is also consistent
with that expected for a Compton-thin AGN and is an order of magnitude
larger than the luminosity ratio for the other heavily obscured AGNs
(see Fig.~3).

The best-fitting parameters for the other heavily obscured AGNs
($\approx$~20--120 net counts; mean of $\approx$~60 net counts) are
determined from jointly fitting the X-ray spectra. Using the absorbed
power-law model, the best-fitting parameters are
$\Gamma=0.35^{+0.43}_{-0.29}$ with $N_{\rm
H}=(5.79^{+12.34}_{-3.90})\times10^{22}$~cm$^{-2}$; see Fig.~4. Such a
flat intrinsic X-ray spectral slope is inconsistent with that found
for typical AGNs and is also inconsistent with that expected from the
inverse Compton scattering of accretion-disk photons (i.e.,\ the X-ray
emitting ``corona''; Haardt \& Maraschi 1993; Mushotzky, Done, \&
Pounds 1993; Reynolds \& Nowak 2003). However, these properties are
consistent with a reflection-dominated spectrum, such as that
typically identified in Compton-thick AGNs and some heavily obscured
Compton-thin AGNs (e.g.,\ George \& Fabian 1991; Matt \etal 1996,
2000; Ueda \etal 2007; Eguchi \etal 2009; Comastri \etal
2010). Indeed, from jointly fitting the X-ray spectra of the heavily
obscured AGNs with a reflection-dominated spectrum (the {\sc pexrav}
model in {\sc xspec}; Magdziarz \& Zdziarski 1995), leaving the
intrinsic X-ray spectral slope as the jointly fitted parameter but
allowing the normalisation of each source to vary, we obtain
$\Gamma=1.69^{+0.15}_{-0.07}$; the reflection parameter in {\sc
pexrav} is fixed to $R=-1$ (to produce only the reflection component),
the cut-off energy is fixed to E\_cut~=~128~keV (e.g.,\ Malizia \etal
2003), and all of the other parameters (inclination angle, elemental
abundances) are fixed at their default values. The best-fitting X-ray
spectral slope is now in good agreement with that found for the
luminous AGNs, providing evidence that these systems are the heavily
obscured reflection-dominated counterparts of the luminous AGNs.

\subsubsection{Reflection-dominated heavily obscured AGNs}

To explore further whether the reflection-dominated model provides a
good description of the X-ray spectra of the heavily obscured AGNs, we
also produced a composite rest-frame \hbox{2--20~keV} spectrum
following \S3.4 of Alexander \etal (2005a). Briefly, the unbinned
spectrum of each object is fitted using a simple power-law model and
an unfolded spectrum is produced, taking into account the {\it
Chandra} effective area and exposure time. Each spectrum is then
converted to rest-frame energies and all of the spectra are combined
and binned to increase the signal-to-noise ratio. The composite X-ray
spectrum of the heavily obscured AGNs is shown in Fig.~5 and is
compared to the {\sc pexrav} model with $\Gamma=1.7$. The similarity
between the composite X-ray spectrum and the pure reflection model is
striking, directly showing that the typical X-ray spectrum of the
heavily obscured AGNs is reflection dominated. The composite X-ray
spectrum is also consistent with that of reflection-dominated AGNs,
such as those recently identified at $z\approx$~0 using $>10$~keV
observatories ({\it Swift}; {\it Suzaku}; e.g.,\ Ueda \etal 2007;
Eguchi \etal 2009; Comastri \etal 2010); see Fig.~5. The similarity
between the composite spectrum of the low-luminosity X-ray sources and
a $\Gamma=2$ power-law spectrum suggests that many of these sources
are either intrinsically weak AGNs (see \S3.1.1 for X-ray variability
constraints) or dominated by HMXBs; see Fig.~2 for the typical range
of X-ray spectral slopes for AGNs and HMXBs.

If the X-ray emission of the heavily obscured AGNs is dominated by
reflection then we would also expect to identify strong Fe~K emission
(e.g.,\ Reynolds \& Nowak 2003). There is no clear evidence for Fe~K
emission in the stacked spectrum but since half of heavily obscured
AGNs have photometric redshifts, the Fe~K emission may be smeared
out. To test this hypothesis we only stacked the X-ray spectra of the
heavily obscured AGNs with spectroscopic redshifts; see inset panel in
Fig.~5. Encouragingly, we now identify a strong emission feature
($\approx$~1~keV rest-frame equivalent width) at $\approx$~6.4~keV,
which is likely to be due to Fe~K emission and suggests the presence
of Compton-thick AGNs, which typically have Fe~K emission with an
equivalent width of $\simgt$~1~keV (e.g., George \& Fabian 1991; Matt
et~al. 1996, 2000; Della Ceca et~al. 2008). However, this feature is
weaker ($\approx$~0.5~keV rest-frame equivalent width) when we remove
J033235.7--274916, which has been individually identified with strong
Fe~K emission (Feruglio et~al. 2011). Since Compton-thin AGNs have
lower equivalent width Fe~K emission than Compton-thick AGNs
($\simlt$~0.5~keV; Mushotzky, Done, \& Pounds 1993; Risaliti
et~al. 2002; Dadina et~al. 2008), the weaker Fe~K emission with
J033235.7--274916 removed suggests that the heavily obscured AGNs
comprise a combination of reflection-dominated Compton-thick and
Compton-thin AGNs (e.g.,\ Matt \etal 2000; Ueda \etal 2007; Eguchi
\etal 2009; Comastri \etal 2010). Under the assumption that the
Compton-thick AGNs have Fe~K equivalent widths of $\simgt$~1~keV, this
suggests that $\simlt$~50\% of the heavily obscured AGNs are absorbed
by Compton-thick material; conversely, a lower limit to the
Compton-thick AGN fraction is $\approx$~10\% due to the identification
of J033235.7--274916 (Feruglio \etal 2011). Qualitatively similar
results have been obtained by Georgakakis et~al. (2010) for IR-excess
galaxies at $z\approx$~1.

Accurate measurements of the intrinsic X-ray luminosity are difficult
in reflection-dominated AGNs due to large uncertainties on the
reflecting geometry. We therefore employ several different approaches
that should bracket the likely range in intrinsic X-ray luminosities.
Clearly, a lower limit is obtained from the {\it observed} 2--8~keV
luminosities, which gives an average luminosity of log($L_{\rm
X}$/erg~s$^{-1}$)$\approx$~43.1 at the average rest-frame energy of
$\approx$~6--24~keV. Conversely, an upper limit is obtained by
assuming that the rest-frame 8~$\mu$m emission is dominated by AGN
activity (i.e.,\ on the basis of the AGN-dominated line in Fig.~3);
following this approach we obtain an average intrinsic X-ray
luminosity of log($L_{\rm 2-10 keV}$/erg~s$^{-1}$)$\approx$~44.0 for
the median rest-frame 8~$\mu$m luminosity of log($L_{\rm 8~\mu
m}/L_{\odot})\approx$~11.1. Lastly, we can estimate the intrinsic
X-ray luminosity of the heavily obscured AGNs under the reasonable
assumption that they are the absorbed counterparts of the luminous
AGNs (see \S3.1.3 for some evidence). Using the average
\hbox{X-ray--8~$\mu$m} luminosity ratio of the luminous AGNs
($\approx$~0.07; a factor $\approx$~3 less than the intrinsic ratio of
$\approx$~0.21), we estimate an average intrinsic luminosity for the
heavily obscured AGNs of log($L_{\rm 2-10
keV}$/erg~s$^{-1}$)$\approx$~43.5. On the basis of this approach we
predict a rest-frame 2--10~keV luminosity within a factor $\approx$~2
of that measured from the optical and X-ray spectroscopy for the
Compton-thick AGN J033235.7--274916 (Feruglio \etal 2011; see Fig.~3),
one of the heavily obscured AGNs in our sample. Since the average
X-ray--8~$\mu$m luminosity ratio of the luminous AGNs is lower than
that expected for AGN-dominated systems, a natural prediction of this
final approach is that the rest-frame 8~$\mu$m emission from many of
the heavily obscured AGNs is dominated by star-formation activity, in
agreement with that found from {\it Spitzer}-IRS spectroscopy (e.g.,\
Murphy et~al. 2009; Donley et~al. 2010; Fadda et~al. 2010).

\subsection{X-ray undetected BzK galaxies}

The majority of the $BzK$ galaxy population remains undetected in the
4~Ms {\it Chandra} observation (see \S3.1) but may host X-ray weak AGN
activity below the detection limit. We can place constraints on the
presence of heavily obscured AGN activity in these systems using X-ray
stacking analyses. We stacked the X-ray data of the X-ray undetected
$BzK$ galaxies adopting the procedure in \S2.5. The X-ray stacking
results are presented in Table~3. As previously found from the X-ray
stacking analyses of $BzK$ galaxies in Daddi \etal (2007a), the result
for the IR-excess galaxies differs from that of the IR-normal
galaxies: $\Gamma=1.4^{+0.3}_{-0.3}$ for the IR-excess galaxies and
$\Gamma=2.0^{+0.4}_{-0.4}$ for the IR-normal galaxies; see
Table~3. The stacked X-ray spectral slope for the IR-normal galaxies
is consistent with that found by Daddi et~al. (2007a) using the 1~Ms
{\it Chandra} data ($\Gamma\approx1.8$) but the X-ray spectral slope
for the IR-excess galaxies is significantly steeper
($\Gamma\approx0.9$ was obtained by Daddi \etal 2007a).

The lack of a flat X-ray spectral slope for the IR-excess galaxies
appears to suggest that we have now individually detected many of the
heavily obscured AGNs that were originally contributing to the stacked
data in Daddi \etal (2007a). However, we must be careful when
interpreting X-ray stacking analyses of X-ray undetected source
populations since the effects of Eddington bias and source variability
(see \S3.1.1) can dominate over the X-ray signal produced by the
majority of the source population. For example, in going from the 1~Ms
{\it Chandra} data to the 4~Ms {\it Chandra} data, the same number of
heavily obscured AGNs were identified in the IR-excess (three of the
13 X-ray detected sources) and IR-normal (three of the 14 X-ray
detected sources) galaxy populations despite there being little
evidence for heavily obscured AGNs in the IR-normal galaxies from the
1~Ms stacked data of Daddi \etal (2007a). It is therefore likely that
further heavily obscured AGNs will be identified with deeper {\it
Chandra} data but we cannot provide direct X-ray constraints from the
current dataset.

%
%
\begin{figure}
\centerline{\includegraphics[angle=0,width=9.0cm]{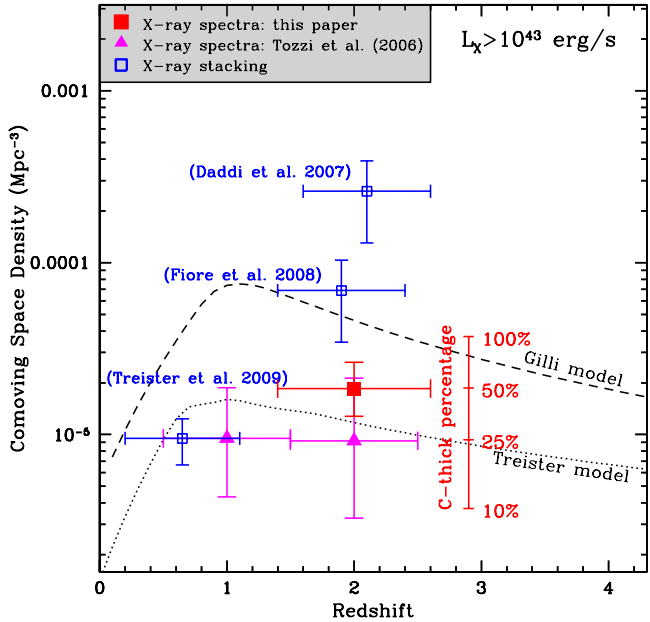}}
\caption{Space density of heavily obscured AGNs with $L_{\rm
  X}\simgt10^{43}$~erg~s$^{-1}$. The plotted data only corresponds to
  the results derived from the X-ray detected heavily obscured AGNs
  identified here (filled squares), candidate Compton-thick AGNs
  identified from X-ray spectral analyses in the CDF-S (filled
  triangles; Tozzi et~al. 2006), and X-ray stacking analysis results
  of X-ray undetected candidate Compton-thick AGNs with $L_{\rm
  X}\simgt10^{43}$~erg~s$^{-1}$ (open squares; Daddi et~al. 2007a;
  Fiore et~al. 2008; Treister et~al. 2009). The solid bar indicates
  the space-density estimates for a range of Compton-thick AGN
  percentages for the heavily obscured AGNs identified here. These
  results are compared to the space-density predictions for
  Compton-thick AGNs with $L_{\rm X}\simgt10^{43}$~erg~s$^{-1}$ based
  on the models of Gilli et~al. (2007; dashed curve) and Triester,
  Urry, \& Virani (2009; dotted curve).}
\end{figure}

\subsection{Re-evaluation of the space density of distant Compton-thick AGNs}

Distant Compton-thick AGNs are of great scientific interest since they
may produce a large fraction of the unresolved $>8$~keV background
(e.g.,\ Worsley et~al. 2005; Gilli et~al. 2007; Treister, Urry, \&
Virani 2009). In \S3.1 we used a variety of analyses (strong
reflected-dominated spectrum, identification of Fe~K, small
X-ray--8~$\mu$m luminosity ratios) to infer that $\approx$~10--50\% of
the X-ray detected heavily obscured AGNs are Compton thick; the lower
limit on the Compton-thick AGN fraction corresponds to the clear
identification of the Compton-thick AGN in J033235.7--274916 (Feruglio
\etal 2011) while the upper limit corresponds to the constraints
derived from the stacked X-ray spectrum when J033235.7--274916 is
removed. We can use these constraints to better estimate the space
density of distant Compton-thick AGNs. In this calculation we have
assumed a broad redshift range of $z=$~1.4-2.6, which gives a comoving
volume of $\approx$~0.7~Gpc$^3$ for the 5.5~arcmin radius region
explored here. We have taken into account the 10\% incompleteness in
the $BzK$ galaxy selection due to blended {\it Spitzer}-IRAC sources
and the 30\% incompleteness due to unreliable UV slopes (see \S5.1 in
Daddi \etal 2007a).

Following the procedure outlined above, we calculate a Compton-thick
AGN space density of $\Phi_{\rm
C-thick}\approx$~$f\times4\times10^{-5}$~Mpc$^{-3}$ at
$z\approx$~1.4--2.6, where f corresponds to the Compton-thick AGN
fraction in our heavily obscured AGN sample; see Fig.~6. As argued in
\S3.1.4, the average intrinsic X-ray luminosity of these heavily
obscured AGNs is $L_{\rm 2-10
keV}\simgt10^{43}$~erg~s$^{-1}$. Although undoubtably uncertain, our
most optimistic space-density estimates lie below the constraints
derived from X-ray stacking analyses of X-ray undetected IR-bright
galaxy populations (Daddi \etal 2007a; Fiore \etal 2008): our space
density estimates are $\simgt$~10 times lower than those of Daddi
\etal (2007a), who used the same object-selection approach as that
adopted here but relied only on X-ray stacking analyses of X-ray
undetected galaxies.  Our space-density constraints are also broadly
consistent with those of Tozzi \etal (2006), who identified
reflection-dominated AGNs using X-ray spectral analyses of sources
detected in the 1~Ms CDF-S observations; we have plotted the Tozzi
\etal (2006) space density in Fig.~5 assuming that 100\% of the
reflection-dominated AGNs are Compton thick and therefore this
represents the maximum space density from that study. However, our
space-density estimate is also a lower limit on the true Compton-thick
AGN space density since (1) there may be further Compton-thick AGNs
with intrinsic $L_{\rm 2-10 keV}\simgt10^{43}$~erg~s$^{-1}$ that lie
below our X-ray detection limit (i.e.,\ those with $N_{\rm
H}\gg10^{24}$~cm$^{-2}$ and a weak reflection component; e.g.,\ Matt
\etal 2000), and (2) our sample does not include distant Compton-thick
AGNs not selected using the $BzK$ technique. An example of the latter
is the $z=1.53$ X-ray bright AGN CXO~J033218.3--275055, which has
strong Fe~K emission identified in the 3~Ms {\it XMM-Newton}
observations of the CDF-S (Comastri \etal 2011) but is not selected as
a $BzK$ galaxy.\footnote{We also note as an aside that only five
($\approx$~45\%) of the 11 X-ray detected heavily obscured AGNs would
be selected using the Fiore \etal (2008) selection criteria of
$R-K>4.5$ and $f_{\rm 24um}/f_R>1000$.} A more complete AGN selection
can be derived using mid-to-far-infrared selection, which is the focus
of a future paper (A.~Del Moro et~al. in prep).

Many studies have predicted the space density of distant Compton-thick
AGNs from the X-ray luminosity functions of relatively unobscured AGNs
and X-ray background constraints (see Ballantyne et~al. 2011 for a
comparison of many of the current studies). In Fig.~6 we compare our
space-density constraints with the predictions from Gilli \etal (2007)
and Treister, Urry, \& Virani (2009) for $L_{\rm
X}\simgt10^{43}$~erg~s$^{-1}$, which broadly represent the most
optimistic and pessimistic estimates, respectively. The model
predictions are already in broad agreement with our range of
space-density measurements, despite our conservative source-selection
approach. However, given the significant uncertainties in the fraction
of Compton-thick AGNs and sample incompleteness, strong conclusions
cannot be derived from the current data.

We can also compare our derived Compton-thick AGN space density to
constraints for other distant AGN populations. From a variety of
studies, the measured space density of $L_{\rm
X}\simgt10^{43}$~erg~s$^{-1}$ AGNs at $z\approx$~2 ranges from
\hbox{$\approx$~(1--2)~$\times10^{-5}$}~Mpc$^{-3}$ for unobscured AGNs
($N_{\rm H}<10^{22}$~cm$^{-2}$) to
$\approx$~\hbox{(1--7)}~$\times10^{-5}$~Mpc$^{-3}$ for all AGNs
(e.g.,\ Barger \etal 2005; Hasinger \etal 2005; La Franca \etal 2005;
Silverman \etal 2008; Aird \etal 2010). On the basis of these studies,
our constraints imply that the space density of Compton-thick AGNs at
$z\approx$~2 is comparable to that of unobscured AGNs at $z\approx$~2
and also suggests that Compton-thick AGNs comprise a non-negligible
fraction of the AGN population at $z\approx$~2. However, our
constraints do not yet support the hypothesis that Compton-thick AGNs
outnumber Compton-thin AGNs at high redshift.  The direct
identification of individual Compton-thick AGN signatures from X-ray,
optical, and mid-IR spectroscopy (e.g.,\ Alexander \etal 2008;
Comastri \etal 2011; Feruglio \etal 2011; Goulding \etal 2011) in a
statistically significant number of objects ($\simgt$~10--20 objects)
are required to provide more reliable constraints.  New facilities
such as {\it NuSTAR} (high-energy 6--78~keV imaging; Harrison \etal
2010) and {\it JWST} (optical--mid-IR spectroscopy; Gardiner \etal
2006) may also provide improved constraints, along with deeper {\it
Chandra}/{\it XMM-Newton} observations and future proposed X-ray
observatories such as {\it WFXT} (Murray \etal 2010).

%
\section{Summary}
%

We have used the 4~Ms {\it Chandra} Deep Field-South observation to
constrain the ubiquity of heavily obscured AGNs in the $z\approx$~2
$BzK$ galaxy population. Our main results are:

\begin{itemize}

\item Forty seven of the 222 $BzK$ galaxies are X-ray detected in the
  central region of the 4~Ms CDF-S field: 11 are heavily obscured AGNs
  ($\Gamma\simlt1$), 9 are luminous AGNs ($L_{\rm
  X}\simgt10^{43}$~erg~s$^{-1}$), and 27 are low-luminosity
  \hbox{X-ray} systems (relatively unobscured AGNs and starburst
  galaxies). Thirteen ($\approx$~48\%) of the 27 X-ray detected $BzK$
  galaxies with reasonable-quality X-ray data (\hbox{$>20$}~counts in
  the 0.5--8~keV band) are found to be variable in the X-ray band,
  including 6 luminous AGNs, 4 heavily obscured AGNs, and 3
  low-luminosity X-ray sources. See \S2.2, \S2.4, and \S3.1.1.

\item The overall X-ray spectra of the heavily obscured AGNs are
  better characterized by a pure reflection model than an absorbed
  power-law model, suggesting extreme Compton-thick absorption
  ($N_{\rm H}\simgt10^{24}$~cm$^{-2}$) in many systems. The
  identification of an emission-line feature at rest-frame
  $\approx$~6.4~keV in the composite \hbox{2--20~keV} spectrum and the
  small X-ray--8~$\mu$m luminosity ratios for the majority of these
  systems provide further support for this interpretation. See
  \S3.1.2--3.1.4.

\item Many of the heavily obscured AGNs are IR-excess galaxies.
  However, only $\approx$~25\% of the X-ray detected IR-excess
  galaxies are heavily obscured AGNs, which is otherwise composed of
  relatively unobscured AGNs and starburst galaxies. See \S3.1.2.

\item X-ray stacking analyses of the X-ray undetected $BzK$ galaxies
  do not clearly reveal the presence of further X-ray undetected AGNs
  below the {\it Chandra} detection limit. This does not rule out the
  possibility that many other heavily obscured AGNs will be detected
  with deeper X-ray observations but it does suggest that they are not
  the dominant X-ray undetected population. See \S2.5 and \S3.2.

\item We estimate a Compton-thick AGN space density of $\Phi_{\rm
  C-thick}\approx$~$f\times4\times10^{-5}$~Mpc$^{-3}$ at
  $z\approx$~1.4--2.6, where $f$ lies between
  $\approx$~0.1--0.5. Although highly uncertain, these constraints are
  already consistent with the range of predictions from X-ray
  background models and imply that the space density of Compton-thick
  AGNs at $z\approx$~2 is comparable to that of unobscured AGNs at
  $z\approx$~2. See \S3.1.4 and \S3.3.

\end{itemize}

\acknowledgements We acknowledge financial support from the Royal
Society (DMA; ACF), a Philip Leverhulme Prize (DMA; JRM), the Science
and Technology Facilities Council (DMA; RCH; ADG; AD), the Chilean
CONICYT grants of FONDECYT 1101024 (FEB) and FONDAP CATA 15010003
(FEB), the {\it Chandra} X-ray Center grants G09-0134B (FEB),
SP1-12007A (WNB; BL; YQX), and SP1-12007B (FEB), the NASA ADP grant
NNX10AC99G (WNB; BL; YQX; MY), the ERC-StG grant UPGAL 240039 (ED),
the French ANR under contract ANR-08-JCJC-008 (ED), and the Italian
Space Agency (ASI) under the ASI-INAF contracts I/009/10/0 and
I/088/06/0 (AC, RG, CV). We thank the referee for a prompt and
considered report and we thank Y.~Ueda for providing the {\it Suzaku}
data and best-fitting model for {\it Swift}~J0601.9-8636 used in
Fig.~5.



%
%

%
%

\clearpage
\LongTables
\begin{landscape}
\begin{deluxetable}{ccccccccccccl}
\tabletypesize{\scriptsize}
\tablewidth{0pt}
\tablecaption{Overall Properties of the X-ray detected $BzK$ Galaxies}
\tablehead{
\multicolumn{2}{c}{$K$-band}        &
\colhead{X--$K$}        &
\colhead{}           &
\colhead{$K_{\rm Vega}$}                  &
\colhead{}           &
\colhead{$L_{\rm UV}$}                  &
\colhead{$L_{\rm 2-10 keV}$}        &
\colhead{$vL_{\rm 8\mu m}$}       &
\colhead{SFR} &
\colhead{} &
\colhead{1~Ms}
\colhead{}\\
\colhead{$\alpha_{J2000}$$^a$}        &
\colhead{$\delta_{J2000}$$^a$}        &
\colhead{(arcsec)$^b$}        &
\colhead{$z$$^c$}           &
\colhead{(mag)$^a$}                  &
\colhead{XID$^d$}                  &
\colhead{log($L_{\odot}$)$^a$}                  &
\colhead{log(erg~s$^{-1}$)$^d$}        &
\colhead{log($L_{\odot}$)$^a$}   &
\colhead{Excess$^a$}       &
\colhead{$\Gamma$$^d$}    &
\colhead{Source?$^e$}    &
\colhead{Notes$^f$}}
\startdata
\\
03 32 10.95 & --27 48 56.1 & 0.48 & 2.81  & 20.2 & 167  &  12.04 &  43.96 &  11.72 &  17.3 &  $1.68^{+0.08}_{-0.09}$ & Y & \\
03 32 12.55 & --27 49 38.2 & 0.29 & 2.45  & 21.3 & 185  &  11.54 &  42.14 &  10.64 &   1.3 &  $>0.53$ & N & \\
03 32 14.12 & --27 49 10.2 & 0.59 & 2.18  & 20.6 & 202  &  11.33 &  42.06 &  11.20 &  15.0 &  $>0.57$ & N & \\
03 32 14.42 & --27 51 10.7 & 0.39 & 1.544 & 19.2 & 205  &  12.11 &  41.98 &  10.72 &   0.5 &  $0.13^{+0.23}_{-0.20}$ & Y & Heavily obscured AGN; optical AGN\\
03 32 14.79 & --27 44 02.5 & 0.67 & 1.56  & 20.7 & 208  &  10.80 &  42.33 &  10.37 &   2.9 &  $1.25^{+0.40}_{-0.29}$ & N & \\
\\
03 32 16.94 & --27 50 04.0 & 0.35 & 1.613 & 20.6 & 236  &  11.40 &  41.59 &  10.50 &   1.1 &  $>0.60$ & N & \\
03 32 17.81 & --27 52 10.3 & 0.57 & 1.76  & 21.7 & 247  &  10.98 & $<$41.95 &  10.47 &   2.7 &  -- & N & \\
03 32 18.24 & --27 52 41.2 & 0.45 & 2.801 & 21.4 & 254  &  11.35 &  43.41 &  10.74 &   2.9 &  $1.79^{+0.26}_{-0.20}$ & Y & Optical AGN\\
03 32 21.30 & --27 51 01.5 & 0.52 & 1.84  & 20.2 & 293  &  11.70 &  41.68 &  11.10 &   4.5 &  $0.36^{+0.64}_{-0.44}$ & N & Heavily obscured AGN\\
03 32 21.99 & --27 51 11.9 & 0.28 & 3.64  & 21.2 & 298  &  11.39 &  42.76 &  11.85 & 124.6 &  $0.01^{+0.21}_{-0.19}$ & Y & Heavily obscured AGN\\
\\
03 32 22.54 & --27 46 03.8 & 0.44 & 1.730 & 20.4 & 308  &  11.26 &  42.90 &  11.00 &   8.7 &  $0.17^{+0.07}_{-0.06}$ & Y & Heavily obscured AGN\\
03 32 22.55 & --27 48 14.9 & 0.48 & 2.54  & 20.4 & 310  &  11.63 &  41.90 &  11.20 &   7.4 & --$0.59^{+0.26}_{0.22}$ & N & Heavily obscured AGN\\
03 32 24.84 & --27 50 50.1 & 0.38 & 2.28  & 20.7 & 337  &  11.47 &  41.91 &  11.14 &   8.7 &  $>0.43$ & N & \\
03 32 25.98 & --27 47 51.3 & 0.69 & 1.90i & 20.5 & 360  &  11.47 &  41.91 &  11.12 &   8.0 &  $>0.94$ & N & IRS: s/burst dominated\\
03 32 28.79 & --27 47 55.5 & 0.95 & 1.383 & 19.3 & 394  &  12.07 &  41.59 &  10.75 &   0.6 &  $>0.64$ & N & \\
\\
03 32 29.09 & --27 46 29.0 & 0.34 & 2.227 & 19.8 & 399  &  11.89 &  41.91 &  11.27 &   5.2 &  $>0.42$ & N & \\
03 32 29.48 & --27 43 22.0 & 0.56 & 1.609 & 19.8 & 405  &  11.66 & $<$41.90 &  10.66 &   1.1 &  $<0.09$ & N & Heavily obscured AGN\\
03 32 29.99 & --27 45 29.9 & 0.38 & 1.218 & 18.3 & 417  &  11.69 &  43.74 &  11.03 &   3.6 &  $1.44^{+0.04}_{-0.04}$ & Y & Optical AGN\\
03 32 31.47 & --27 46 23.2 & 0.43 & 2.223 & 19.0 & 435  &  12.41 &  42.43 &  11.71 &   7.2 &  $1.00^{+0.32}_{-0.27}$ & Y & Heavily obscured AGN\\
03 32 31.52 & --27 48 53.8 & 0.43 & 1.879 & 20.3 & 437  &  11.66 &  42.41 &  11.25 &   8.3 &  $1.31^{+0.34}_{-0.27}$ & Y & IRS: s/burst dominated\\
\\
03 32 31.55 & --27 50 28.6 & 0.76 & 1.613 & 18.9 & 436  &  12.14 & $<$41.64 &  11.01 &   1.2 &  $<0.70$ & N & Heavily obscured AGN\\
03 32 32.93 & --27 50 40.5 & 0.82 & 2.50  & 20.5 & 451  &  11.96 &  41.92 &  11.08 &   2.3 &  $>0.23$ & N & \\
03 32 34.04 & --27 50 28.7 & 0.44 & 1.384 & 19.5 & 467  &  11.76 &  41.42 &  10.71 &   1.0 &  $>0.46$ & N & \\
03 32 34.46 & --27 50 04.9 & 0.21 & 2.14  & 20.5 & 474  &  11.46 &  41.99 &  11.41 &  22.5 &  $>0.77$ & N & \\
03 32 34.98 & --27 49 31.9 & 0.97 & 2.55  & 21.9 & 481  &  11.44 &  42.03 &  11.30 &  16.1 &  $>0.51$ & Y & \\
\\
03 32 35.72 & --27 49 16.1 & 0.31 & 2.578 & 20.0 & 490  &  11.85 &  42.51 &  12.25 & 171.0 &  $0.44^{+0.23}_{-0.20}$ & Y & Heavily obscured AGN; optical AGN\\
03 32 35.97 & --27 48 50.4 & 0.31 & 1.309 & 19.1 & 493  &  11.93 &  42.24 &  10.55 &   0.4 &  $1.98^{+0.43}_{-0.34}$ & Y & \\
03 32 36.17 & --27 51 26.5 & 0.37 & 1.613 & 20.0 & 499  &  11.69 &  42.67 &  10.80 &   1.6 &  $1.27^{+0.18}_{0.16}$  & Y & \\
03 32 36.18 & --27 46 27.6 & 0.43 & 2.48  & 20.9 & 501  &  11.96 &  41.99 &  11.03 &   1.9 &  $>0.41$ & N & \\
03 32 37.36 & --27 46 45.5 & 0.62 & 1.843 & 20.0 & 512  &  11.71 &  41.74 &  11.00 &   3.0 &  $>0.42$ & N & \\
\\
03 32 37.74 & --27 50 00.6 & 0.81 & 1.619 & 19.2 & 517  &  12.23 &  41.71 &  11.16 &   1.6 &  $>0.64$ & N & \\
03 32 37.77 & --27 52 12.3 & 0.27 & 1.603 & 18.8 & 518  &  11.98 &  44.16 &  11.64 &  15.1 &  $1.71^{+0.03}_{-0.03}$ & Y & Optical AGN; IRS: AGN dominated\\
03 32 37.96 & --27 53 07.9 & 0.26 & 1.97  & 21.2 & 520  &  11.44 &  42.60 &  10.55 &   1.2 &  $>1.41$ & Y & \\
03 32 38.55 & --27 46 34.2 & 0.77 & 2.55i & 21.3 & 525  &  11.68 &  42.27 &  11.77 &  46.7 &  $>0.86$ & N & IRS: s/burst dominated\\
03 32 39.08 & --27 46 02.1 & 0.17 & 1.216 & 18.3 & 537  &  11.85 &  43.22 &  10.82 &   1.2 &  $1.13^{+0.05}_{-0.05}$ & Y & Optical AGN\\
\\
03 32 39.74 & --27 46 11.5 & 0.08 & 1.552 & 19.4 & 549  &  11.63 &  43.24 &  11.08 &   4.9 &  $1.16^{+0.07}_{-0.07}$ & Y & IRS: s/burst dominated\\
03 32 40.06 & --27 47 55.4 & 0.45 & 1.998 & 19.5 & 552  &  12.08 &  41.99 &  11.58 &   9.8 &  $>0.63$ & N & IRS: s/burst dominated\\
03 32 40.76 & --27 49 26.2 & 0.40 & 2.130 & 19.7 & 555  &  11.85 &  42.05 &  11.34 &   7.2 &  $>0.65$ & N & IRS: s/burst dominated\\
03 32 41.80 & --27 51 35.3 & 0.20 & 1.63  & 20.2 & 562  &  11.54 &  41.90 &  10.94 &   3.7 &  $0.77^{+0.57}_{-0.38}$ & N & Heavily obscured AGN\\
03 32 43.25 & --27 49 14.3 & 0.20 & 1.920 & 19.5 & 577  &  11.31 &  43.96 &  11.10 &  10.9 &  $1.55^{+0.04}_{-0.04}$ & Y & Optical AGN\\
\\
03 32 43.46 & --27 49 01.8 & 0.46 & 1.78i & 20.2 & 579  &  11.76 &  42.00 &  11.27 &   7.0 &  $>0.88$ & Y & IRS: s/burst dominated\\
03 32 43.61 & --27 46 59.0 & 0.20 & 1.570 & 20.4 & 580  &  11.46 &  41.86 &  10.66 &   1.7 &  $>0.83$ & N & \\
03 32 44.02 & --27 46 34.9 & 1.00 & 2.688 & 20.9 & 583  &  11.75 &  43.63 &  11.69 &  31.1 &  $2.00^{+0.15}_{-0.16}$ & Y & IRS: s/burst dominated\\
03 32 44.37 & --27 49 11.3 & 0.53 & 2.13  & 20.4 & 589  &  11.25 & $<$42.03 &  11.44 &  40.5 &  $>0.60$ & N & \\
03 32 44.60 & --27 48 36.0 & 0.18 & 2.593 & 21.4 & 593  &  11.55 &  43.38 &  11.11 &   6.6 &  $1.19^{+0.12}_{-0.11}$ & Y & \\
\\
03 32 46.84 & --27 51 20.9 & 0.25 & 2.292 & 20.4 & 617  &  11.74 &  42.42 &  11.06 &   3.6 &  $>0.73$ & N & \\
03 32 47.72 & --27 50 38.0 & 0.40 & 1.63  & 19.2 & 625  &  12.06 &  41.94 &  11.21 &   2.8 & --$0.33^{+0.20}_{-0.19}$ & N & Heavily obscured AGN\\
\enddata
\tablenotetext{a}{$BzK$ galaxy properties taken from Daddi \etal
(2007b). Co-ordinates correspond to the $K$-band position of the $BzK$
galaxy. The UV luminosity corresponds to rest-frame 1500~\AA\ and has
been corrected for extinction (see \S3.6 of Daddi \etal 2007b). The
rest-frame 8~$\mu$m luminosity is calculated using the 24~$\mu$m flux
density, with small K-corrections applied (see \S3.1 of Daddi \etal
2007b). The SFR excess corresponds to the ratio of star-formation
rates (mid-IR+UV versus extinction-corrected UV; see \S2.2 of Daddi
\etal 2007a).}
\tablenotetext{b}{Offset between the position of the X-ray source and
the $K$-band $BzK$ counterpart in arcseconds.}
\tablenotetext{c}{$BzK$ galaxy redshifts. Optical spectroscopic
redshifts are denoted by having three decimal places and come from
Szokoly et~al. (2004), Mignoli et~al. (2005), Cimatti et~al. (2008),
Vanzella et~al. (2008), Popesso et~al. (2009), and Balestra
et~al. (2010). All other redshifts are either photometric (from
Grazian \etal 2006; Luo et~al. 2010; Rafferty et~al. 2011) or from
{\it Spitzer}-IRS spectroscopy (highlighted with ``i'' and from
Teplitz et~al. 2007 and Fadda et~al. 2010).}
\tablenotetext{d}{X-ray source properties. XID corresponds to the
X-ray identification number in Xue et~al. (2011). The X-ray spectral
slope ($\Gamma$) determined from the band ratio (2--8~keV to
0.5--2~keV count-rate ratio) plus 1~$\sigma$ uncertainty are taken
from Xue \etal (2011); $\Gamma$ is re-calculated from the band ratio
for the low-count sources listed in Xue \etal (2011) with
$\Gamma=1.4$. The rest-frame 2--10~keV luminosity is calculated from
the observed-frame 0.5--2~keV flux in Xue et~al. (2011) and converted
to rest-frame 2--10~keV assuming $\Gamma=1.8$.}
\tablenotetext{e}{Indicates if the source was detected in the 1~Ms
  CDF-S catalogs of Alexander et~al. (2003); otherwise the source was
  detected in the 4~Ms CDF-S catalogs of Xue et~al. (2011).}
\tablenotetext{f}{Notes and classifications. Objects classified as
``Heavily obscured AGN'' have $\Gamma\simlt1$ and objects classified
as optical AGN have optical spectroscopic signatures from Szokoly
\etal (2004) indicating AGN activity. We have also indicated which
sources have {\it Spitzer}-IRS spectroscopy, available either from
Teplitz \etal (2007), Donley \etal (2010), or Fadda \etal (2010).}
\end{deluxetable}
\clearpage
\end{landscape}

%
%

\clearpage
\LongTables
\begin{deluxetable}{cccccccc}
\tabletypesize{\scriptsize}
\tablecolumns{8} 
\tablewidth{0pt} 
\tablecaption{X-ray Properties of the X-ray detected $BzK$ Galaxies}
\tablehead{
\multicolumn{2}{c}{X-ray}        &
\colhead{}	&
\multicolumn{4}{c}{X-ray variability constraints\tablenotemark{c}}        &
\colhead{Variable}\\
\colhead{$\alpha_{J2000}$$^a$}								&
\colhead{$\delta_{J2000}$$^a$}	&
\colhead{Net counts\tablenotemark{b}}	&
\colhead{$\chi^2$} 								 	&
\colhead{$P_{\chi^2}$}							 	&
\colhead{$\sigma_{rms}^2$}							&
\colhead{$\sigma_{\text{HMXB}}^2$}					&
\colhead{AGN?\tablenotemark{d}}}
\startdata
\\
03 32 10.98 & --27 48 56.5  &  844    &   4.09		&  0.004  &   0.014  &  0.0054  &  Y 	 \\
03 32 12.57 & --27 49 38.2  &  9      &   \dots		&  \dots  &   \dots  &  \dots  &  \dots  	\\
03 32 14.15 & --27 49 10.6  &  18     &   \dots	&  \dots  &  \dots  &  \dots  &  \dots  	\\
03 32 14.43 & --27 51 11.0  &  82     &   4.30		&  0.005  &   0.141  &  0.0011  &  Y  	\\
03 32 14.80 & --27 44 03.2  &  89     &   0.371	&  0.723  &  -0.034  &  N/A  &  N  	\\
\\
03 32 16.96 & --27 50 04.2  &  5      &   \dots	&  \dots  & \dots  &  \dots  &  \dots   \\
03 32 17.84 & --27 52 10.8  &  27     &   0.952	&  0.289  &  -0.001  &  0.0030  &  N  	\\
03 32 18.24 & --27 52 41.6  &  215    &   1.46		&  0.183  &   0.009  &  0.0019  &  N   \\
03 32 21.31 & --27 51 02.0  &  22     &   0.688	&  0.416  &  -0.092  &  0.0110  &  N   \\
03 32 22.00 & --27 51 12.1  &  80     &   0.106	&  0.952  &  -0.058  &  0.0110  &  N   \\
\\
03 32 22.56 & --27 46 04.2  &  564    &   6.47		&  0.000  &   0.036  &  0.0005  &  Y   \\
03 32 22.58 & --27 48 15.2  &  58     &   0.228	&  0.842  &  -0.073  &  0.0110  &  N   \\
03 32 24.85 & --27 50 50.4  &  7      &   \dots	&  \dots  &  \dots  &  \dots  &  \dots  \\
03 32 26.01 & --27 47 51.8  &  8      &   \dots		&  \dots  &  \dots  &  \dots  &  \dots   \\
03 32 28.85 & --27 47 56.0  &  17     &   \dots		&  \dots  &  \dots  &  \dots  &  \dots   \\
\\
03 32 29.11 & --27 46 29.3  &  12     &   \dots	&  \dots  &  \dots  &  \dots  &  \dots   \\
03 32 29.50 & --27 43 22.5  &  54     &   0.682	&  0.451  &  -0.038  &  N/A  &  N   \\
03 32 29.99 & --27 45 30.3  &  3916   &   130 	&  0.000  &   0.163  &  0.0003  &  Y   \\
03 32 31.48 & --27 46 23.6  &  59     &   0.870	&  0.352  &  -0.017  &  0.0000  &  N   \\
03 32 31.55 & --27 48 54.0  &  61     &   0.670	&  0.481  &  -0.046  &  0.0081  &  N   \\
\\
03 32 31.51 & --27 50 29.0  &  16     &   \dots	&  \dots  &  \dots   &  \dots  &  \dots   \\
03 32 32.95 & --27 50 41.3  &  10     &   \dots	&  \dots  &  \dots   &  \dots  &  \dots   \\
03 32 34.03 & --27 50 29.1  &  12     &   \dots  	&  \dots  &  \dots   &  \dots  &  \dots   \\
03 32 34.47 & --27 50 05.1  &  10     &   \dots  	&  \dots  &  \dots   &  \dots  &  \dots   \\
03 32 35.04 & --27 49 32.5  &  4      &   \dots  	&  \dots  &  \dots   &  \dots  &  \dots   \\
\\
03 32 35.72 & --27 49 16.4  &  70     &   1.64		&  0.144  &   0.019  &  0.0110  &  N   \\
03 32 35.98 & --27 48 50.7  &  84     &   5.26		&  0.001  &   0.300  &  0.0014  &  Y   \\
03 32 36.18 & --27 51 26.8  &  179    &   2.13		&  0.067  &   0.027  &  0.0009  &  N   \\
03 32 36.19 & --27 46 28.0  &  3      &   \dots	&  \dots  &  \dots   &  \dots  &  \dots   \\
03 32 37.38 & --27 46 46.1  &  11     &   \dots	&  \dots  &  \dots   &  \dots  &  \dots   \\
\\
03 32 37.75 & --27 50 01.4  &  14     &   \dots	&  \dots  &  \dots   &  \dots  &  \dots   \\
03 32 37.77 & --27 52 12.6  &  4600   &   250		&  0.000  &   0.021  &  0.0001  &  Y   \\
03 32 37.96 & --27 53 08.2  &  71     &   3.38	  	&  0.014  &   0.141  &  0.0004  &  Y   \\
03 32 38.52 & --27 46 34.9  &  14     &   \dots	&  \dots  &  \dots   &  \dots  &  \dots   \\
03 32 39.09 & --27 46 02.1  &  1331   &   134		&  0.000  &   0.346  & 	0.0005  &  Y   \\
\\
03 32 39.74 & --27 46 11.5  &  760    &   0.833	&  0.453  &  -0.001  &  0.0003  &  N   \\
03 32 40.05 & --27 47 55.8  &  26     &   1.50		&  0.145  &  -0.014  &  0.0110  &  N   \\
03 32 40.77 & --27 49 26.6  &  13     &   \dots	&  \dots  &  \dots   &  \dots  &  \dots   \\
03 32 41.82 & --27 51 35.4  &  37     &   4.09		&  0.007  &   0.508  &  0.0007  &  Y   \\
03 32 43.24 & --27 49 14.5  &  2117   &   39.2		&  0.000  &   0.077  &  0.0004  &  Y   \\
\\
03 32 43.45 & --27 49 02.2  &  19     &   \dots		&  \dots  &   \dots  &  \dots  &  \dots   \\
03 32 43.61 & --27 46 59.2  &  16     &   \dots	&  \dots  &  \dots   &  \dots  &  \dots   \\
03 32 44.04 & --27 46 35.9  &  380    &   1.13		&  0.281  &   0.001  &  N/A  &  N   \\
03 32 44.41 & --27 49 11.3  &  7      &   \dots	&  \dots  &  \dots   &  \dots  &  \dots   \\
03 32 44.61 & --27 48 36.2  &  333    &   3.40		&  0.011  &   0.033  &  0.0003  &  Y   \\
\\
03 32 46.86 & --27 51 21.0  &  43     &   4.34		&  0.006  &   0.335  &  0.0056  &  Y  	\\
03 32 47.73 & --27 50 38.4  &  118    &   10.4		&  0.000  &   0.210  &  0.0002  &  Y   \\
\enddata
\tablenotetext{a}{Co-ordinates correspond to the X-ray position from
Xue et~al. (2011).}
\tablenotetext{b}{Background-subtracted (net) counts in the 0.5-8.0
keV band used in the X-ray spectral analyses and X-ray variability
analyses.}
\tablenotetext{c}{X-ray variability constraints for the sources with
reasonable-quality X-ray data ($>20$~counts in the 0.5--8~keV band). The $\chi^2$
statistic is calculated by comparing the variability observed between
observations to that expected from Poisson statistics. $P_{\chi^2}$
gives the probability that the observed variability is due to Poisson
noise, $\sigma_{rms}^2$ gives the normalized excess variance (Nandra
et~al. 1997), and $\sigma_{\text{HMXB}}^2$ gives the upper limit to
the HMXB contribution to the variability. N/A indicates that no SFR
information was available and so no attempt was made to estimate
$\sigma_{\text{HMXB}}^2$: this does not affect any significantly
variable sources. See \S2.4.}
\tablenotetext{d}{Indicates if the source was found to show excess
X-ray variability over that expected from the HMXB population; see
\S2.4.}
\end{deluxetable}

%
%

\clearpage
\LongTables
\begin{deluxetable}{lccccccccc}
\tabletypesize{\scriptsize}
\tablewidth{0pt}
\tablecaption{X-ray Stacking Analyses of the X-ray Undetected $BzK$ Galaxies}
\tablehead{
\colhead{}        &
\colhead{}           &
\colhead{}    &
\colhead{$L_{\rm UV}$}                  &
\colhead{$vL_{\rm 8\mu m}$} &
\colhead{0.5--2 keV}              &
\colhead{2--8 keV}              &
\colhead{}      &
\colhead{}              &
\colhead{$L_{\rm 2-10 keV}$}\\
\colhead{Sample}        &
\colhead{$N$}           &
\colhead{$z$$^a$}              &
\colhead{log($L_{\odot}$)$^a$}                  &
\colhead{log($L_{\odot}$)$^a$} &
\colhead{(10$^{-6}$~cts~s$^{-1}$)$^b$}              &
\colhead{(10$^{-6}$~cts~s$^{-1}$)$^b$}              &
\colhead{Band ratio$^c$}    &
\colhead{$\Gamma$$^c$}      &
\colhead{log(erg~s$^{-1}$)$^d$}}
\startdata
& & & & & & & & &\\
IR excess galaxies: $K<22$         & 47 & $1.97\pm0.34$ & $11.45\pm0.46$ & $11.01\pm0.33$ & 1.47 (12.6)  & 0.84 (3.9) & $0.57\pm0.15$ & $1.4\pm0.3$ & 41.40  \\
IR normal galaxies: $K<22$         & 116& $1.93\pm0.44$ & $11.48\pm0.26$ & $10.53\pm0.30$ & 1.19 (15.8)  & 0.42 (2.8) & $0.33\pm0.12$ & $2.0\pm0.4$ & 41.32  \\
\vspace{0.05in} 
\tablenotetext{a}{Median galaxy properties and MAD (see Footnote 3):
redshift, exinction-corrected UV luminosity (rest-frame 1500~\AA; see
\S3.6 of Daddi \etal 2007b), rest-frame 8~$\mu$m luminosity
(calculated using the 24~$\mu$m flux density, with small K-corrections
applied; see \S3.1 of Daddi \etal 2007b).}  \tablenotetext{b}{Count
rates in the 0.5--2~keV and 2--8~keV bands; the numbers in parantheses
correspond to the S/N ratio.}  \tablenotetext{c}{X-ray spectral
properties: band ratio (2--8~keV to 0.5--2~keV count-rate ratio) and
X-ray spectral slope ($\Gamma$), derived from the band ratio, and
1~$\sigma$ uncertainties. Calculated following Xue et~al. (2011).}
\tablenotetext{d}{Rest-frame 2--10~keV luminosity calculated from the
0.5--2~keV flux and converted to rest-frame 2--10~keV assuming
$\Gamma=1.8$; the 0.5--2~keV flux is calculated following Xue \etal
(2011).}
\end{deluxetable}

\end{document}